\documentclass{appolb}
\usepackage{epsfig}

\usepackage{graphicx}
\usepackage{amsmath}
\usepackage{array}
\usepackage{feynmf}

\def\lg{{\mathchoice{~\raise.58ex\hbox{$<$}\mkern-14.8mu\lower.52ex\hbox{$>$}~}
                    {~\raise.58ex\hbox{$<$}\mkern-14.8mu\lower.52ex\hbox{$>$}~}
                    {\raise.59ex\hbox{{$\scriptscriptstyle <$}}\mkern-12.8mu%
                     \lower.01ex\hbox{{$\scriptscriptstyle >$}}}   {}   }}
\def\gl{{\mathchoice{~\raise.58ex\hbox{$>$}\mkern-12.8mu\lower.52ex\hbox{$<$}~}
                    {~\raise.58ex\hbox{$>$}\mkern-12.8mu\lower.52ex\hbox{$<$}~}
                    {\raise.62ex\hbox{{$\scriptscriptstyle >$}}\mkern-12.0mu%
                     \lower.05ex\hbox{{$\scriptscriptstyle <$}}}  {}    }}

\newcommand{\be}{\begin{equation}}
\newcommand{\ee}{\end{equation}}
\newcommand{\ba}{\begin{eqnarray}}
\newcommand{\ea}{\end{eqnarray}}
\newcommand{\ban}{\begin{eqnarray*}}
\newcommand{\ean}{\end{eqnarray*}}
\newcommand \nn {\nonumber}
\newcommand{\sla}{\!\!\!/ \,}

\begin{document}

\title{Super Yang-Mills Plasma\thanks{Presented by St. Mr\' owczy\' nski at XXXI Max Born Symposium and HIC for FAIR Workshop {\it Three Days of Critical Behaviour in Hot and Dense QCD}, Wroc\l aw, Poland, June 14-16,  2013}}

\author{Alina Czajka
\address{Institute of Physics, Jan Kochanowski University, Kielce, Poland}
\and
Stanis\l aw Mr\' owczy\' nski
\address{Institute of Physics, Jan Kochanowski University, Kielce, Poland \\
National Centre for Nuclear Research, Warsaw, Poland}
}

\maketitle
\begin{abstract}
The ${\cal N} =4$ super Yang-Mills plasma is studied in the regime of weak coupling. Collective excitations and collisional processes are discussed and compared to those of QCD plasma. The two systems are concluded to be very similar to each other with the differences mostly reflecting different numbers of degrees of freedom. 
\end{abstract}

\PACS{52.27.Ny, 11.30.Pb, 03.70.+k}

  
\section{Introduction}

A great interest in the ${\cal N} =4$ super Yang-Mills theory, which is conformally invariant not only at the classical but at the quantum level as well,  was stimulated by a discovery of the AdS/CFT duality of the five-dimensional gravity in the anti de Sitter geometry and the conformal field theories \cite{Maldacena:1997re}. The duality offered a unique tool to study strongly coupled field theories. Since the gravitational constant and the coupling constant of dual conformal field theory are inversely proportional to each other, some problems of strongly coupled field theories can be solved via weakly coupled gravity. In this way some intriguing features of strongly coupled systems driven by the ${\cal N} =4$ super Yang-Mills dynamics were revealed, see the reviews \cite{Son:2007vk,Janik:2010we}. However,  one asks how properties of the ${\cal N} =4$ super Yang-Mills plasma (SYMP) are related to those of quark-gluon plasma (QGP) studied experimentally in relativistic heavy-ion collisions. Some properties of strongly coupled SYMP have been confronted with those of QGP, see {\it e.g.} \cite{Panero:2009tv}, but, in general, such a comparison is a difficult problem. Instead some comparative analyses have been done in the domain of weak coupling where perturbative methods are applicable \cite{CaronHuot:2006te,Huot:2006ys,CaronHuot:2008uh,Blaizot:2006tk,Chesler:2006gr,Chesler:2009yg}.

We undertook a task of systematic comparison of supersymmetric perturbative plasmas to their non-supersymmetric counterparts. We started with the ${\cal N} =1$ SUSY QED, analyzing first collective excitations of ultrarelativistic plasma which, in general, is out of equilibrium \cite{Czajka:2010zh} and then, in the subsequent paper \cite{Czajka:2011zn} we discussed collisional characteristics. Our findings show that the SUSY QED and QED plasmas are surprisingly similar to each other.  Further we have studied the ${\cal N} =4$ super Yang-Mills plasma, analyzing again collective excitations and collisional characteristics \cite{Czajka:2012gq}. Here we summarize the study.

Throughout the paper we use the natural system of units with $c= \hbar = k_B =1$; our choice of the metric tensor is $(+ - - -)$.

\section{${\cal N}=4$ Super Yang-Mills Theory}
\label{sec-lagrangian}

The gauge group is ${\rm SU}(N_c)$ and every field belongs to its  adjoint representation. There are gauge bosons (gluons) described by the vector field $A_\mu^a$ with $a, b, c, \dots = 1, 2, \dots N_c^2 -1$. There are four Majorana fermions represented by the Weyl spinors combined in the Dirac bispinors $\Psi_i$ with $i = 1,2,3,4$. Finally, there are six real scalar fields which are assembled in the multiplet $\Phi = (X_1, Y_1, X_2, Y_2, X_3, Y_3)$, where $X_p$ and $Y_p$ are scalars and pseudoscalars. The Lagrangian can be written as \cite{Yamada:2006rx}
\ba
\label{Lagrangian-1}
{\cal L}
&=&
-\frac{1}{4}F^{\mu \nu}_a F_{\mu \nu}^a
+\frac{i}{2}\bar \Psi_i^a (D\!\sla \Psi_i)^a
+\frac{1}{2}(D_\mu \Phi_A)_a (D^\mu \Phi_A)_a
\\ [2mm] \nn
&&
-\frac{1}{4} g^2f^{abe} f^{cde} \Phi_A^a \Phi_B^b \Phi_A^c \Phi_B^d
-i\frac{g}{2} f^{abc} \Big( \bar \Psi_i^a  \alpha_{ij}^p  X_p^b \Psi_j^c  
+i\bar \Psi_i^a \beta_{ij}^p\gamma_5  Y_p^b \Psi_j^c \Big),
\ea
where
$F^{\mu \nu}_a = \partial^\mu A^\nu_a - \partial^\nu A^\mu_a + g f^{abc} A^\mu_b A^\nu_c$ and the covariant derivatives equal $(D\!\sla \Psi_i)^a = (\partial \,\!\sla \delta_{ab} +g f^{abc} A_c \!\! \sla) \Psi_i^b$ and $(D^\mu \Phi)_a = D^\mu_{ab} \Phi_b = (\partial^\mu \delta_{ab} + gf^{abc}A^\mu_c)\Phi_b$; $g$ is the coupling constant; $f^{abc}$ are the structure constants of ${\rm SU}(N_c)$ group; the $4 \times 4$ matrices $\alpha^p, \beta^p$ satisfy the commutation relations $\{\alpha^p, \alpha^q \} = - 2 \delta^{p q}$, $\{\beta^p, \beta^q \} = - 2 \delta^{p q}$, $[ \alpha^p, \beta^q] = 0 $. Their explicit form is given  in \cite{Yamada:2006rx}.

\section{Basic plasma characteristics}
\label{sec-basic}

As in QGP, there are several conserved charges in SYMP. Comparing the two systems we assume that all average charges and the associated chemical potentials vanish.  Then, the temperature $(T)$ is the only dimensional parameter which characterizes the equilibrium plasma. Taking into account the right numbers of bosonic and fermionic degrees of freedom in SYMP and QGP, the energy densities of equilibrium non-interacting plasmas equal
\be
\varepsilon = \frac{\pi^2}{60} \, {30 (N_c^2 -1)  \choose 4(N_c^2 -1) + 7 N_f N_c}  T^4.
\ee
where the upper expression is for SYMP and the lower one for QGP with $N_f$ light quark flavors. For $N_c=N_f =3$, the energy density of SYMP is approximately 2.5 times bigger than that of QGP at the same temperature. The same holds for the pressure $p$ which, obviously, equals $\varepsilon/3$.

The Debye masses in SYMP and QGP equal 
\be
m_D^2 = \frac{g^2}{6} \, {12 N_c \choose  2 N_c  + N_f}   T^2.
\ee
For $N_c=N_f =3$, the ratio of Debye masses squared is 2.4 at the same value of $gT$. The Debye mass determines not only the screening length $r_D = 1/m_D$ but it also gives the plasma frequency  $\omega_p = m_D/\sqrt{3}$ which is the minimal frequency of longitudinal and transverse plasma oscillations.

Another important quantity characterizing the equilibrium plasma is the so-called plasma parameter $\lambda$ which equals the inverse number of particles in the sphere of radius of the screening length. When $\lambda$ is decreasing, the behavior of plasma is more and more collective while  inter-particle collisions are less and less important. For $N_c=N_f =3$, we have
\be
\lambda \equiv \frac{1}{\frac{4}{3} \pi r_D^3 n} \approx {0.257 \choose 0.042} g^3 .
\ee
The dynamics of QGP is thus more collective. The difference of energy densities of SYMP and of QGP merely reflects the difference in numbers of degrees of freedom. For $m_D$ and $\lambda$ it also matters that fermions in QGP and SYMP  belong to different representations of the ${\rm SU}(N_c)$ group. 

\section{Dispersion equations and self-energies}
\label{sec-dis-eqs}

Knowing the field equations of motion, one writes down the gluon, fermion and scalar dispersion equations as
\be
\label{dis-photon-1}
{\rm det}\big[ k^2 g^{\mu \nu} -k^{\mu} k^{\nu} - \Pi^{\mu \nu}(k) \big]
 = 0 
\ee
\be
\label{dis-electron-1}
 {\rm det}\big[ k\sla  - \Sigma (k) \big]  = 0,
~~~~~~~~~~~~
k^2  + P(k)   = 0 ,
\ee
where color and other indices are dropped, $\Pi^{\mu \nu}(k)$, $\Sigma (k)$ and $P(k)$ are the retarded self energies and $k\equiv (\omega, {\bf k})$ is the four-momentum. As seen, the whole dynamical information is contained in the self-energies.

To compute the self-energies, which enter the dispersion equations, the plasma is assumed to be homogeneous, locally colorless but the momentum distribution is, in general, different from equilibrium one. Therefore, the Keldysh-Schwinger formalism, which allows one to describe both equilibrium and non-equilibrium many-body systems, is adopted. We also apply the Hard Loop Approach, see the review  \cite{Kraemmer:2003gd}, which was generalized to anisotropic systems in \cite{Mrowczynski:2000ed}. 

Computing the one-loop contributions and performing the Hard-Loop Approximation, one finds
the retarded gluon polarization tensor as
\be
\label{Pi-k-final}
\Pi^{\mu \nu}_{ab}(k)
= g^2 N_c \delta_{ab}
\int \frac{d^3p}{(2\pi)^3}
\frac{f({\bf p})}{E_p} 
\frac{k^2 p^\mu p^\nu - (k^\mu p^\nu + p^\mu k^\nu - g^{\mu \nu} (k\cdot p))
(k\cdot p)}{(k\cdot p + i 0^+)^2},
\ee
where 
\be
\label{f-def}
f({\bf p}) \equiv 2n_g({\bf p}) + 8n_f({\bf p}) + 6n_s({\bf p})
\ee
is the effective distribution function of plasma constituents. The coefficients in front of the distributions functions $n_g({\bf p})$,   $n_f({\bf p})$, $n_s({\bf p})$ equal the numbers of degrees of freedom (except colors) of, respectively, gauge bosons, fermions and scalars. This is a manifestation of supersymmetry. Another effect of the supersymmetry is vanishing of the tensor (\ref{Pi-k-final}) in the vacuum limit when $f({\bf p})  = 0$. The polarization tensor (\ref{Pi-k-final}) is symmetric $(\Pi^{\mu \nu}(k) = \Pi^{\nu \mu}(k))$ and transverse $(k_\mu \Pi^{\mu \nu}(k) = 0)$ and thus it is gauge independent.

The fermion and scalar self-energies computed at the one-loop level in the Hard-Loop Approximation are
\ba
\label{Si-k-final}
\Sigma^{ij}_{ab}(k) &=& \frac{g^2}{2} \, N_c \delta_{ab}\delta^{ij}
\int \frac{d^3p}{(2\pi )^3}
\frac{f({\bf p})}{E_p}  \, \frac{p\sla}{k\cdot p + i 0^+},
\\[2mm]
\label{P-k-final}
P^{AB}_{ab}(k) &=& - 2g^2 N_c \delta_{ab} \delta^{AB}
\int \frac{d^3p}{(2\pi)^3} \frac{f({\bf p})}{E_p},
\ea
and, as the polarization tensor (\ref{Pi-k-final}), they depend on the function (\ref{f-def}).

\section{Effective Action}
\label{sec-eff-action}

The Hard Loop Approach can be formulated in an elegant and compact way by introducing the effective action which was first derived for equilibrium plasmas in \cite{Taylor:1990ia,Frenkel:1991ts,Braaten:1991gm} and later on generalized to anisotropic systems in \cite{Pisarski:1997cp,Mrowczynski:2004kv}.

Since the self-energy of a given field is the second functional derivative of the action with respect to the field, one writes
\be
\label{action-A-1}
{\cal L}^{(A_\mu^a)}_2(x) =
\frac{1}{2} \int d^4y \; A_\mu^a(x) \Pi_{ab}^{\mu \nu}(x-y) A_\nu^b(y) ,
\ee
where $\Pi_{ab}^{\mu \nu}$ is given by the formula (\ref{Pi-k-final}). The subscript `2' indicates that the action generates only two-point functions. To get $n$-point functions the action needs to be modified to a gauge invariant form: the ordinary derivative should be replaced by the covariant one. Repeating the calculations described in \cite{Mrowczynski:2004kv}, one finds the Hard Loop effective actions as
\ba
\label{action-A-2}
{\cal L}^{A}_{\rm HL} &=& g^2 N_c  \int \frac{d^3p}{(2\pi )^3} \,
\frac{f({\bf p})}{E_p} \,
F_{\mu \nu}^a (x) \bigg({p^\nu p^\rho \over (p \cdot D)^2}\bigg)_{ab} F_\rho^{b \;\mu} (x) ,
\\ [2mm]
\label{action-Psi-2}
{\cal L}^{\Psi}_{\rm HL} &=& g^2 N_c
\int \frac{d^3p}{(2\pi )^3} \, \frac{ f({\bf p})}{E_p} \,
\bar{\Psi}^a_i(x) \bigg( {p \cdot \gamma \over p\cdot D}\bigg)_{ab} \Psi^b_i(x) ,
\\ [2mm]
\label{action-Phi-2}
{\cal L}^{\Phi}_{\rm HL} &=& - 2g^2 N_c
\int \frac{d^3p}{(2\pi )^3} \, \frac{f({\bf p})}{E_p} \;
\Phi_A^a(x) \Phi_A^a(x) .
\ea
where $f({\bf p})$ is, as previously, the  distribution function (\ref{f-def}).

The actions  (\ref{action-A-2}, \ref{action-Psi-2}, \ref{action-Phi-2}) are obtained from the self-energies but the reasoning can be turned around. As argued in \cite{Frenkel:1991ts,Braaten:1991gm}, the actions of gauge bosons (\ref{action-A-2}), fermions (\ref{action-Psi-2}), and scalars (\ref{action-Phi-2}) are of unique gauge invariant form. Therefore, the structures of hard-loop self-energies are unique. Consequently, the self-energies can be inferred from the known QED and QCD results with some help of supersymmetry arguments. 

\section{Collective modes}
\label{sec-modes}

When the self-energies are substituted into the dispersion equations, collective modes are found as solutions of the equations. 

The structure of polarization tensor (\ref{Pi-k-final}) is such as of gluon polarization tensor in QCD plasma. It also has  analogical form as in both usual and supersymmetric QED plasma. Therefore, the spectrum of collective excitations of gauge bosons is in all cases the same. In equilibrium plasma we have the longitudinal and transverse plasmons. When the plasma is out of equilibrium there is a whole variety of possible collective excitations. In particular, there are unstable modes, see  {\it e.g.} the review  \cite{Mrowczynski:2007hb}, which exponentially grow in time and strongly influence the system's dynamics.

The form of Majorana fermion self-energy (\ref{Si-k-final}) happens to be the same as the quark self-energy in QCD plasma. It also coincides with the electron self-energy in both non-supersymmetric and supersymmetric QED plasma. Therefore, we have identical spectrum of excitations of fermions in all these systems. In equilibrium plasma there are two modes of opposite helicity over chirality ratio.  In non-equilibrium plasma the spectrum of fermion collective excitations changes but no unstable modes have been found even for an extremely anisotropic momentum distribution  \cite{Mrowczynski:2001az,Schenke:2006fz}.

The scalar self-energy (\ref{P-k-final}) is  independent of momentum, it is negative and real. Therefore,  one writes $P(k) = - m^2_{\rm eff}$ where $m_{\rm eff}$ is the effective mass. Then, the dispersion equation is solved when $k^2 = m^2_{\rm eff}$.

\section{Collisional characteristics}
\label{sec-collisions}

Transport coefficients of weakly coupled QGP, which include baryon and strangeness diffusion, electric charge and heat conductivity, shear and bulk viscosity and color conductivity, have been studied in detail, see \cite{Arnold:2000dr,Arnold:2003zc,Arnold:2006fz,Arnold:1998cy} and references therein. The shear viscosity of SYMP has been computed in \cite{Huot:2006ys} and the bulk viscosity  is identically zero because of exact conformality. Other transport coefficients of SYMP have not been studied but they are expected to be qualitatively similar to those of QGP.

Let us consider, for example, the shear viscosity $\eta$. Since the temperature is the only dimensional parameter, which characterizes the equilibrium plasma of massless constituents, $\eta$ must be proportional to $T^3$. The dominant contributions to $\eta$ of QGP come from the binary collisions driven by a one-gluon exchange. The analysis presented in \cite{Arnold:2000dr} shows that at the leading order $\eta \sim T^3/g^4\ln g^{-1}$. The factor $1/\ln g^{-1}$ appears due to the infrared singularity of the Coulomb-like interaction. 

One expects the same parametric form of $\eta$ and other transport coefficients in case of SYMP and QGP because, similarly to QGP, there are the Coulomb-like binary interactions for every constituent of SYMP. The analysis \cite{Huot:2006ys} indeed proves that the shear viscosity coefficients of QGP and SYMP differ only by numerical factors which mostly reflect different numbers of degrees of freedom in the two plasmas. The viscosity is strongly dominated by the Coulomb-like interactions, and it does not much matter that the sets of elementary processes in the two plasmas are different. 

We considered \cite{Czajka:2011zn} two transport characteristics of the ${\cal N} =1$ QED plasma which are not so constrained by dimensional arguments and seemed to strongly depend on elementary process under consideration. Specifically, we computed the collisional energy loss and momentum broadening of a particle traversing the equilibrium plasma.  The dimensional argument does not work here because the two quantities depend not only on the plasma temperature but on the energy of test particle as well. We computed the energy loss and momentum broadening due to the processes which, like the Compton scattering on selectrons, are independent of momentum transfer. Such processes are qualitatively different from the Coulomb-like interactions dominated by small momentum transfers.  We managed to obtain the exact formulas of the energy loss and momentum broadening due to the momentum-independent scattering. In the limit of high energy of test particle, which is important in the context of jet suppression phenomenology, the energy loss and momentum broadening appeared to be very similar (at the leading order) to those driven by the Coulomb-like interactions. 

The result can be understood as follows. One estimates the energy loss  $\frac{dE}{d x}$ as $\langle \Delta E \rangle / \lambda$, where $\langle \Delta E \rangle$ is the typical change of particle's energy in a single collision and $\lambda$ is the particle's mean free path given as $\lambda^{-1} = \rho \, \sigma$ with $\rho \sim T^3$ being the density of scatterers and $\sigma$ denoting the cross section. For the differential cross section, which is independent of momentum transfer, the total cross section is $\sigma \sim e^4/s$. When a highly energetic particle with energy $E$ scatters on massless plasma particle, $s \sim ET$ and consequently  $\sigma \sim e^4/(ET)$. The inverse mean free path is thus estimated as $\lambda^{-1} \sim e^4 T^2/E$.  When the scattering process is independent of momentum transfer, $\langle \Delta E \rangle$ is of order $E$ and we finally find $-\frac{dE}{d x} \sim e^4 T^2$. In case of Coulomb interaction we have $\langle \Delta E \rangle \sim - e^2 T$, $\lambda^{-1} = e^2 T$ which provide the same estimate of the energy loss. The energy transfer in a single collision is thus much smaller in the Coulomb interaction than in the momentum independent scattering but the cross section is bigger in the same proportion.  Consequently, the two interactions corresponding to very different differential cross sections lead to very similar energy losses. 

We expect an analogous situation in SYMP. There are various elementary process but the energy loss and momentum broadening of highly energetic particles do not much differ from those in QGP.

\section{Conclusions}
\label{sec-conclusions}

QCD is obviously rather different than ${\cal N} = 4$ super Yang-Mills theory. Nevertheless QGP and SYMP are surprisingly similar in the weak coupling regime (at the leading order). The form of gluon collective excitations is identical and the same is true for the fermion (quark) modes.  The scalar modes in SYMP are as of massive relativistic particle. The sets of elementary processes are different in QGP and SYMP but the transport coefficients, which are dominated by the Coulomb-like interactions, are quite similar.  The energy loss and momentum broadening of a highly energetic test particle are also rather similar in the two plasma systems. The differences mostly come from different numbers of degrees of freedom in both plasmas which need to be taken into account for a quantitative comparison. 

\vspace{4mm}

This work was partially supported by the Polish National Science Centre under Grant No. 2011/03/B/ST2/00110.


\end{document}